\documentclass[9pt,conference]{IEEEtran}
\IEEEoverridecommandlockouts

\usepackage[absolute,showboxes]{textpos}
\TPMargin{5pt}
\newcommand{\copyrightstatement}{
\begin{textblock}{0.8}(0.1,0.01)
\noindent
\footnotesize
\copyright 2025 IEEE.  Personal use of this material is permitted.  Permission from IEEE must be obtained for all other uses, in any current or future media, including reprinting/republishing this material for advertising or promotional purposes, creating new collective works, for resale or redistribution to servers or lists, or reuse of any copyrighted component of this work in other works.
\end{textblock}
}

\usepackage{cite}
\usepackage{amsmath,amssymb,amsfonts}
\usepackage{algorithmic}
\usepackage{graphicx}
\usepackage{textcomp}
\usepackage{xcolor}
\usepackage{url}

\def\tt{{\tilde{t}}}

\def\BibTeX{{\rm B\kern-.05em{\sc i\kern-.025em b}\kern-.08em
    T\kern-.1667em\lower.7ex\hbox{E}\kern-.125emX}}
\begin{document}
\copyrightstatement

\title{30+ Years of Source Separation Research: \\Achievements and Future Challenges}

\author{
    \IEEEauthorblockN{Shoko Araki\textsuperscript{1}, Nobutaka Ito\textsuperscript{2}, Reinhold Haeb-Umbach\textsuperscript{3}, Gordon Wichern\textsuperscript{4}, Zhong-Qiu Wang\textsuperscript{5}, Yuki Mitsufuji\textsuperscript{6}}
    \IEEEauthorblockA{
        \textsuperscript{1}NTT Corporation, Japan, \textsuperscript{2}University of Tokyo, Japan, \textsuperscript{3}Paderborn University, Germany, \\ \textsuperscript{4}Mitsubishi Electric Research Laboratories, USA, \textsuperscript{5}Southern University of Science and Technology, China, \textsuperscript{6}Sony AI, USA
    }
}


\maketitle

\begin{abstract}
Source separation (SS) of acoustic signals is a research field that 
emerged in the mid-1990s and has flourished ever since. On the occasion of ICASSP's 50\textsuperscript{th} anniversary, we review the major contributions and advancements in the past three decades in the speech, \textcolor{black}{audio}, and music SS research field.
We will cover both single- and multi-channel SS approaches.
We will also look back on key efforts to foster a culture of scientific evaluation in the research field, including challenges, performance metrics, and datasets. We will conclude by discussing current trends and future research directions.
\end{abstract}


\section{Introduction}
\label{sec:intro}

Source separation (SS) is a technology that reconstructs the individual source signals from one or more mixtures of them. 
Given that daily acoustic environments usually contain multiple concurrent sound sources,
SS is an important technology for acoustic signals, e.g., listening to each speaker's voice when multiple people speak simultaneously or extracting vocal or specific instrumental parts from music. 

SS, including blind SS (BSS), of acoustic signals~\cite{BSS,ASSE,ASS}  
is an important area of research that rapidly emerged from around the mid-1990s to the early 2000s and 
much research is still being conducted in this field.
The SS topic was added to the EDICS in 2006 as AUD-SSEN (Source Separation and Signal Enhancement), 
and then 
separated into AUD-SEP (Audio and Speech Source Separation) and AUD-SEN (Signal Enhancement and Restoration) to differentiate between those growing research areas in 2014. 
In recent years, ICASSP has consistently received 40--50 submissions to AUD-SEP every year. 
Over the past 30+ years, many researchers have entered the field and many breakthroughs have been made.
In addition, since the 2000s, when BSS research based on independent component analysis (ICA) was being conducted, various attempts were made to make the SS field prosperous as described in Sec.~\ref{sec:initiatives}. Furthermore, since deep learning was introduced in audio SS in the 2010s, many data-driven methods have been proposed, and a lot of SS research has been conducted not only in the audio and acoustic signal processing (AASP) community but also in the speech and machine learning communities.

This paper will review not only the key contributions and advancements in the last three decades, but also key efforts such as challenges, performance metrics, and datasets to foster SS research. We will also discuss current trends and future directions in SS research.

\section{Problem description}
Given $N\,(\geq 2)$ source signals $s_n(\tt)$, where $\tt$ is the discrete time index, the mixture signal $y_m(\tt)$ captured at the $m^{\text{th}}$ of  $M\,(\geq 1)$ microphones, can be written as a convolutive mixture
\begin{align}
    y_m(\tt) &= \sum_{n=1}^{N}\sum_{\tau} h_{mn}(\tau)s_{n}(\tt-\tau) + v_m(\tt) 
     \end{align}
    \begin{align} 
    &= \sum_{n=1}^{N}c_{mn}(\tt) + v_m(\tt)~~~ \text{with}\,\,m \in \{1,\cdots,M\},
\label{eq:obs_t}
\end{align}
where $h_{mn}$ is the acoustic impulse response from source $n$ to microphone $m$, and $v_m$ is an additive noise term, which is often neglected.

The goal of SS is to estimate the source signals $s_n$ or their images $c_{\mu n}$ at a reference microphone $\mu$, from the observed mixtures $\{y_1,\dots,y_M\}$.
If this problem is solved without leveraging any prior knowledge about the sources or mixing conditions, it is referred to as {\it blind} SS (BSS).
The cases where $N<M$, $N=M$, and $N>M$ are called over-determined, determined, and underdetermined problems, respectively. 
In many studies, the number of sources $N$ is
assumed known, even for BSS. 

Under certain assumptions,  the convolutive mixture can be approximated as an instantaneous mixture in the time-frequency (TF) domain by using, e.g., the short-time Fourier transform (STFT):
\begin{equation}
        y_{mtf} = \sum_{n=1}^N h_{mnf}s_{ntf} + v_{mtf},
\end{equation}
where $t$ and $f$ are time frame and frequency bin indices, respectively. While this is often more computationally tractable, it introduces a frequency permutation problem in narrow-band SS algorithms, since the order in which the source signals appear at the separator output at different frequencies is arbitrary.

\section{Source separation approaches}
This section introduces representative SS technologies that have been proposed over the past 30+ years. 

\subsection{Model-based SS
}
BSS is inherently an ill-posed problem. Thus additional assumptions must be made to arrive at a unique solution. 
Assumptions on the source signals and/or the mixing process are given in mathematical models to solve BSS. Approaches to BSS can be distinguished by the assumptions applied.  

\vspace*{1mm}
\noindent{\bf Multi-channel approaches:}

For determined or over-determined BSS, {\bf independent component analysis (ICA)} was applied to separate acoustic signals in the mid-1990s \cite{Bell1995}. This approach assumes that the source signals are non-Gaussian, non-stationary, non-white, or a combination of those.

Methods for dealing with convolutive mixtures were actively studied in the TF domain (e.g., \cite{BSS,ASSE,ASS}). 
To address the permutation problem in frequency-domain ICA, {\bf independent vector analysis (IVA)} \cite{IVE_Taesu,IVA_Hiroe,Ono2011auxIVA} was proposed to find a separating matrix based on the independence between vectors bundling signal components of all frequencies, assuming that frequency components of the same source are statistically dependent. 
The permutation indeterminacy can also be addressed by employing a full-band source model. A popular choice is {\bf non-negative matrix factorization (NMF)}, which models a spectrogram as linear combinations of base spectra. The combination of frequency-domain ICA and NMF as a source model is known as {\bf independent low-rank matrix analysis (ILRMA)}~\cite{Kitamura2016}. 
Solving the permutation problem in frequency-domain ICA and accelerating algorithms are still important research topics.


For an underdetermined problem ($N>M$), {\bf TF masking} is one of the primary methods. By utilizing W-disjoint orthogonality \cite{DUET} of many sound source signals in the TF domain, this method extracts the dominant sound at each TF with a binary (or soft) mask. A typical mask estimation method is clustering spatial features at each TF obtained from multi-channel mixtures in an expectation-maximization (EM) framework using a spatial mixture model (e.g.,~\cite{cACGMM}). Recently, many mask estimation approaches using a neural network have been proposed, as described in Sec.~\ref{subsec:DL}. 
The source signals can then be extracted by simply multiplying the microphone signal in the TF domain with the estimated masks. 
Alternatively, the TF masks are used to estimate beamformer coefficients, leading to source extraction by beamforming, which typically reduces artifacts~\cite{Souden,Yoshioka_Chime3,GSS}.

Another approach to underdetermined BSS is
{\bf full-rank spatial covariance analysis (FCA)}~\cite{FCA}. FCA is capable of handling situations in which the W-disjoint orthogonality does not hold, such as reverberant environments and music SS. To this end, the method uses a multi-channel Wiener filter instead of the TF mask and necessitates spatial covariance matrices of the individual sources.
These matrices are modeled as full-rank instead of rank-one to better deal with reverberant environments and are estimated from the observed mixtures using the EM algorithm.
Key advancements include avoiding the permutation problem by incorporating NMF~\cite{Arberet2010,Sawada2013},
boosting SS performance by combining with a deep generative model~\cite{Bando2021}, and
reducing computation through joint diagonalization~\cite{Ito2021,Sekiguchi2020}.

In music separation, methods that work well in the underdetermined case are required since music is predominantly delivered in a stereo format and typically contains more than two instruments (i.e., sources). Panning is often used in music production such that sources have different locations in the stereo image, and by assuming that there is very little overlap between different sources in the magnitude spectrogram (i.e., the W-disjoint orthogonality), histograms of angle estimates for each time-frequency bin can be used to create separation masks. DUET~\cite{rickard2007duet}, ADRess~\cite{barry2004sound} and PROJET~\cite{fitzgerald2016projet} are well-known examples of energy vs angle separation algorithms, which, while imperfect due to the W-disjoint orthogonality assumption, are often lightweight enough to run in real-time.


\vspace*{1mm}
\noindent{\bf Single-channel approaches:}

SS has to rely on spectral cues to discriminate sources if only a single-channel input is available. An early attempt to solve this extremely hard problem was {\bf factorial hidden Markov models (HMMs)}, where a multi-dimensional dynamic programming algorithm was employed to track and decode the time trajectory of the individual sources \cite{HERSHEY2010}. However, the grammar was very restrictive, and general multi-talker separation was far from being solved. 

Assuming source magnitudes are additive, {\bf NMF} \cite{LeeSeung1999} decomposes the magnitude spectrum of the mixture into a product of a basis matrix of prototype spectra and an activation matrix. NMF guarantees the non-negativity of the estimated base spectra and activations, making it well suited for magnitude spectra.
NMF has been predominantly applied to audio and music SS, where the source spectra of individual sounds or instruments can have quite different signatures \cite{Virtanen2007}.

The introduction of neural networks, which learn the spectral patterns of the source signals in a supervised learning phase brought about a real breakthrough, even for single-channel input, as described in the next section.

In music separation, specific qualities of music signals can be exploited for designing single-channel separation algorithms. REpeating Pattern Extraction Technique (REPET)~\cite{rafii2013repeating} identifies periodically repeating segments in music signals, which typically correspond to background musical elements, and can then be separated from non-repeating foreground elements, such as lead vocals. Another computationally efficient approach is harmonic-percussive separation~\cite{fitzgerald2010harmonic}, which uses median filtering on both the time and frequency dimensions of the magnitude spectrogram to separate harmonic elements (approximately horizontal spectrogram lines) from percussive elements (vertical lines). If a musical score is available, using it as a hint has also been an active research area~\cite{ewert2014score}.

\subsection{Deep learning-based SS
}
\label{subsec:DL}

For audio SS,  the sound types
(e.g., speech, music, and sound events) 
to deal with are often known as a prior. 
One can leverage any prior knowledge of signal patterns to address the ill-posed problem in SS, thereby improving separation.
In recent years, deep learning has shown remarkable capability at learning signal patterns from massive data.
In this context, \textit{learning to separate} based on supervised learning and deep neural networks (DNNs) has attracted broad research interest and become a promising
direction \cite{Wang2017overview}.

A major breakthrough in this direction was made by Wang and Wang \cite{Y.Wang2013}.
Using simulated pairs of clean and noisy speech, they realized speech enhancement by training DNNs on spectral features to predict the so-called {\bf ideal binary mask}.
At run time, the estimated mask 
functions like a Wiener filter that can suppress noise.
This concept formulates speech enhancement as a data-driven classification problem and can be approached via large-scale supervised learning on massive simulated data.

Unlike speech enhancement, where speech and noise sources 
exhibit different signal patterns, speaker separation has a particular difficulty in label permutation, since the sources to separate are all speech  and homogeneous.
This poses a major challenge for supervised learning, as the sources estimated by DNNs are often not aligned with the true sources.
A major breakthrough that addresses this issue is {\bf deep clustering} \cite{Hershey2016}, which trains DNNs to embed each TF unit so that the embeddings of the TF units dominated by the same source are close to each other and far away otherwise. At run time, binary TF masks for separating sources are estimated by clustering the learned embeddings.
Another key approach to address the issue is {\bf permutation invariant training (PIT)} \cite{Hershey2016, Yu2017a}, which aligns estimated sources with true sources before loss computation.

Another major direction
is {\bf target speaker extraction}, which informs DNNs about which sources to separate by inputting auxiliary information,
such as speaker embeddings or visual cues~\cite{ Ephrat2018Look2Listen,TSE}.

Even in monaural conditions, where only spectral cues can be utilized for SS, supervised deep learning has already shown remarkable effectiveness \cite{Wang2017overview}.
When multiple microphones are available, 
spatial features can complement
spectral ones to improve separation \cite{Wang2018iSpatialAndSpectralFeatures}.
Since in many applications the same microphone array is used in training and testing, a trend is to directly stack the real and the imaginary components~\cite{Wang2021MCCSM} or, more simply, the waveforms~\cite{Liu2020WaveformMapping} of input mixtures as input features. %
In scenarios where the array geometry could be different between training and testing, DNN modules that can model arrays with various geometries and numbers of microphones have been proposed \cite{Luo2020GeometryInvariant, Yoshioka2022}.

In the past decade, DNN-based SS approaches have shifted from TF masking, which estimates only the target magnitude via real-valued masking and uses the mixture phase for signal re-synthesis, to complex spectrum estimation~\cite{Williamson2016CRM, Tan2020CSM}, or even to time-domain waveform estimation~\cite{Luo2019ConvTasNet}.
%
This shift has been propelled by the rapid development of deep learning.
In early work, only feed-forward DNNs with fully-connected layers were utilized \cite{Y.Wang2013}.
Convolutional neural networks, which can better model local signal patterns, were later leveraged \cite{Stoller2018WaveUNet}.
Subsequently, recurrent neural networks with long short-term memory (LSTM) were introduced to model temporal patterns \cite{Erdogan2015}.
Recently, Transformers with attention mechanisms were designed to model long-term signal dependencies for SS \cite{Subakan2021SepFormer}.
Modern DNN architectures in SS often combine various DNN blocks to leverage their complementary power.
For example, the popular convolution recurrent network \cite{Tan2018CRN} sandwiches an LSTM with a U-Net so that local and longer-term signal patterns can be integrated for separation.
There are also fully convolutional networks such as {\bf Conv-TasNet} \cite{Luo2019ConvTasNet}, which operates on short frames in the time domain.
Inspired by the seminal {\bf dual-path recurrent neural networks (DPRNN)}~\cite{Luo2020DPRNN}, state-of-the-art DNN architectures in SS usually employ a dual- or multi-path architecture, where signal patterns are alternately modeled along different tensor axes.
For example, in {\bf TF-GridNet} \cite{Wang2023TFGridNet}, 
a sub-band temporal module and an intra-frame full-band module are alternately stacked to model the 
temporal and spatial patterns within each frequency and the spectral and spatial patterns in each frame.

SS is not limited to speech
but can be applied to various types of audio signals.
One such application is music, which often includes vocals and instruments.
The first significant success was achieved with Open-Unmix~\cite{stoter19}, which employs LSTM and data augmentation for stereo music \cite{UhlichPGEKTM17}.
Later, combining hybrid architectures \cite{TakahashiGM18}, using different data representations \cite{défossez2022}, and bridging multiple instrument networks \cite{SawataTUTM24} showed improved performance. 
Another application is mixed sound events.
A seminal work is universal sound separation \cite{Kavalerov2019USS}, which trained an improved Conv-TasNet via supervised PIT or unsupervised {\bf mixture invariant training} \cite{Wisdom2020MixIT} to unmix mixed sound events.
Another direction is target sound extraction, where the auxiliary cues informing the DNN can be a binary vector indicating a subset of a pre-defined set of sound events to separate \cite{Ochiai2020} or natural language which can offer more flexible prompting \cite{Liu2023SeparateAnything}.


\subsection{Hybrid SS}\label{hybrid_SS_main_body}

The classical way to exploit spatial information present in multi-channel input is to use a beamformer that points a beam of increased sensitivity towards the source of interest. 
The issue with SS through beamforming is that the beamformer weight computation requires knowledge of the statistics of the source signals, which, naturally, is not readily available if only the mixture is given. A popular way to obtain this information is through TF mask estimation, whereby the spatial covariance matrix of a source is estimated from the TF bins that the mask estimator has classified to be dominated by that source. TF mask estimation can be done either by clustering spatial features as described earlier or by using a DNN that learns the spectro-temporal patterns of the source signals. The latter, which employs a DNN for parameter estimation and beamformers for source extraction, is an example of a hybrid approach, that blends signal processing with deep learning. This hybrid method was first introduced in the CHiME-$3$ challenge, where it was extremely successful \cite{Heymann_2015}.

Hybrid techniques can also be used for mask estimation itself, because spatial clustering and DNNs exploit different signal properties: the former takes advantage of spatial and the latter of spectral information. The former relies on unsupervised,  while the latter on supervised learning. These complementary properties have been exploited in various ways, e.g., for overcoming domain mismatch between training and test \cite{nakatani2017integrating} or for training a DNN for SS on real mixtures, where the separated signals as a training target are not available \cite{Drude_2019, Tzinis_2019}. 

\section{Initiatives that fostered SS research
}
\label{sec:initiatives}

In the 2000s, there were no benchmarks in SS research, and researchers often used their own datasets and evaluation criteria. 
This made it difficult to discuss the advantages and disadvantages of technologies and to make objective comparisons, and thus a common language and common base were needed in the community.
A number of efforts have been made to remedy this situation and foster SS research,
including challenges, performance metrics, datasets, and open-source software. 
This section introduces these key efforts. 

The International Conference on Independent Component Analysis and Blind Signal Separation (renamed the International Conference on Latent Variable Analysis and Signal Separation in 2010), held every 1.5 years since 1999, has also played an important role in developing the fundamental theory of SS and its applications to audio. 


\subsection{Challenges}\label{ssec:challenges}

In conjunction with the conference, Signal Separation Evaluation Campaign (SiSEC) was launched in 2007 to benchmark various separation models and continued until  2018 \cite{SiSEC2}.
The SiSEC addressed the above issue by providing a dataset, evaluation metrics, and sample software codes. 
SiSEC originated from the community-based Signal Separation Evaluation Campaign (SASSEC)\footnote{\url{https://www.irisa.fr/metiss/SASSEC07/}}, which was organized in 2007 and provided development and test sets for speech and music. 
Since then, it has been organized in $2008$, $2010$, $2011$, $2013$, $2015$, $2016$, and $2018$ (see references in \cite{SiSEC1,SiSEC2}).
Although the data size of the SiSECs was small compared to today, it created a culture of benchmarking and challenges in the SS research field.

This SiSEC initiative was taken over by CHiME\footnote{\url{https://www.chimechallenge.org/}}. The first PASCAL CHiME challenge (CHiME-1) in 2010 was on separating and recognizing speech in everyday listening conditions. Since then, CHiME has continued to provide challenge projects using data recorded in real environments and to lead the SS field.

Subsequently, the music separation~(MUS) track of SiSEC was succeeded by a crowd-based competition called the Music Demixing~(MDX) Challenge $2021$ \cite{Mitsufuji22}. This was followed by an expanded edition, the Sound Demixing~(SDX) Challenge $2023$, featuring both music and cinematic demixing tracks~\cite{Fabbro24, Uhlich24}.

\subsection{Evaluation metrics}


Vincent {\it et al.}~\cite{BSSEval} introduced a well-known set of evaluation metrics, encompassing the source-to-distortion ratio (SDR), source-to-interference ratio (SIR), and sources-to-artifacts ratio (SAR). 
These metrics are based on decomposing each separated signal into components corresponding to the target sound source, residual interference from the other sources, and artifacts introduced during separation, such as musical noise. 
Implementations are available in the widely-used BSS Eval Matlab toolbox\footnote{\url{https://gitlab.inria.fr/bass-db/bss_eval/}} and in Python libraries\footnote{\url{https://craffel.github.io/mir_eval/}}\footnote{\url{https://github.com/sigsep/sigsep-mus-eval/}}.
Additionally, a modified metric known as the scale-invariant SDR (SI-SDR)~\cite{LeRoux2019} was proposed, which allows only a rescaling factor, rather than a finite impulse response filter, to align the target source signal with each separated signal.
The word error rate (WER) is often used to measure the effectiveness of SS as an automatic speech recognition (ASR) front-end. 
Short-time objective intelligibility (STOI)~\cite{STOI1} is an objective measure of speech intelligibility based on the correlation coefficient between the short-time temporal envelopes of the target and the separated speech. 
The mean opinion score (MOS) measures perceptual quality using human subjective evaluation scores collected in a listening test. Since obtaining subjective scores is time-consuming and costly, objective metrics, such as perceptual evaluation of speech quality (PESQ)~\cite{PESQ3}, are often used to predict perceptual quality.

\subsection{Datasets and open source}

As deep learning approaches began to dominate SS research, publicly available datasets became increasingly necessary to enable reproducible scientific research. Particularly important examples include CHiME and MUSDB18~\cite{SiSEC2}\footnote{https://zenodo.org/records/1117372}, which served as the basis for the challenges described in Sec.~\ref{ssec:challenges}. Because SS training requires both source signals and mixtures, scripts that combine audio signals from existing datasets (e.g., speech datasets originally collected for ASR) further aided reproducibility. For example, speech SS research was facilitated by scripts for creating the wsj0-2mix dataset~\cite{Hershey2016}, which was later extended to WHAMR!~\cite{whamr} in noisy and reverberant scenarios and to LibriMix~\cite{librimix} with proprietary speech data replaced with open data.
SMS-WSJ~\cite{smswsj} enabled evaluation as an ASR front-end and that of multi-channel SS methods.
With the recent release of the EARS dataset~\cite{richter2024ears}, there now exists a dataset with high sampling rate speech recordings in anechoic conditions, which should further advance the field.

In addition to publicly available datasets, the proliferation of the open source software ecosystem has aided SS research as it has in many other fields. Notable tools include pyroomacoustics~\cite{scheibler2018pyroomacoustics} for simulating audio sources in reverberant environments, gpurir~\cite{diaz2021gpurir} for enabling reverberation modeling on the GPU, and Nara-WPE~\cite{Drude_2019} for providing one of the first open implementations of a state-of-the-art dereverberation algorithm. For deep learning algorithms, the open-sourcing of training recipes and pre-trained models has been the key to accelerating research, such as Open-Unmix~\cite{stoter19}, Asteroid~\cite{Pariente2020Asteroid}, and ESPnet-SE~\cite{li2021espnet}.


\section{Remaining challenges and potential directions}

SS performance on many simulated benchmarks under constrained setups (e.g., fully overlapped speech signals, a given number of sources) has saturated. The research community has been moving toward more and more realistic conditions, such as real-recorded conversational speech in CHiME-7 and 8. DNN-based SS algorithms have shown limited success in such real-world settings, where there remain two main challenges.
First, simulated data used for supervised DNN training are usually mismatched with real data, which may cause poor generalization to real data. To alleviate this, recent efforts in unsupervised, weakly supervised, and semi-supervised SS~\cite{Drude_2019,Tzinis_2019,Wisdom2020MixIT,Huang2022,Wang2024SuperM2M} aim to leverage real (``in-the-wild") data without ground-truth sources. Generative models, such as diffusion models~\cite{DiffSep}, may also handle out-of-domain data. 
Second, separating an unknown and time-varying number of sources necessitates source activity detection (e.g., speaker diarization, sound event detection) and needs further exploration.

Other challenges include improving the performance of SS with a single or a limited number of microphones.
Additionally, the development of evaluation metrics applicable to real-world mixtures (i.e., reference-free) that strongly correlate with human perception for not only speech signals, but also music and general sounds remains challenging.
Moreover, as in other areas, developing lightweight SS models for 
edge devices and pursuing low-latency algorithms for real-time applications continue to be or become increasingly important issues.
Given that sound sources often move in applications such as robot audition,  multi-channel SS for moving sources remains a significant challenge.
Another promising avenue is the synchronization of observed mixtures across independent devices to form a distributed microphone array. We look forward to further development of SS research in the following decades.


\vspace*{2mm}
\noindent{\bf Acknowledgement} We would like to thank Dr.~Mike Goodwin for 
providing the EDICS history of SS. 


\bibliographystyle{IEEEtran}
\bibliography{refs}

\end{document}